\documentclass[aps,twocolumn,prl,superscriptaddress,amsmath,amssymb,amsfonts]{revtex4-1} 

\usepackage[T1]{fontenc}
\usepackage[utf8]{inputenc}
\usepackage{lmodern} % load a font with all the characters
\usepackage{color}
\usepackage{bbold}
\usepackage{dsfont}
\usepackage{graphicx}
\usepackage[caption=false]{subfig}
\usepackage[colorlinks]{hyperref}
\usepackage{physics}
\usepackage{amsmath}
\usepackage{amssymb}
\usepackage{array} 
\usepackage{xspace}
\usepackage{soul}
\usepackage{xparse}
\usepackage{physics}
\usepackage{siunitx}

\usepackage{multibib}

\renewcommand{\ket}[1]{\ensuremath{\left|#1\right\rangle}}

\def\OD{\text{OD}\xspace}

\newcommand{\MHz}{\text{MHz}}

\def\deltar{\delta_s}
\def\chibar{\bar{\chi}}
\def\FGR{Fermi's Golden Rule\xspace}

\def\be{\begin{equation}}
\def\ee{\end{equation}}

\def\beqa{\begin{eqnarray}}
\def\eeqa{\end{eqnarray}}
\newcommand{\beq}{\begin{equation} \vspace{-0em}} 
\newcommand{\eeq}{\vspace{-0.em} \end{equation}}
\newcommand{\integral}[1]{\int \! \mathrm{d} #1\,}                    % integral without limits
\def\nn{\nonumber}
\newcommand{\ra}{\ensuremath{\rightarrow}\xspace}
\newcommand{\rs}{\rm \scriptscriptstyle}

\newcommand{\figref}[1]{Fig.~\ref{#1}}

\def\mid{D}

\renewcommand{\eqref}[1]{Eq.~\ref{#1}}
\def\VoneD{\tilde{V}}
\def\Ve{\ensuremath{V_e}\xspace}

\begin{document}

\title{Tunable three-body loss in a nonlinear Rydberg medium}% Force line breaks with \\

\author{Dalia P. Ornelas-Huerta}
\affiliation{Joint Quantum Institute, NIST/University of Maryland, College Park, Maryland 20742 USA}

\author{Przemyslaw Bienias}
\affiliation{Joint Quantum Institute, NIST/University of Maryland, College Park, Maryland 20742 USA}
\affiliation{Joint Center for Quantum Information and Computer Science, NIST/University of Maryland, College Park, Maryland 20742 USA}

\author{Alexander N. Craddock}
\affiliation{Joint Quantum Institute, NIST/University of Maryland, College Park, Maryland 20742 USA}

\author{Michael J. Gullans}
\affiliation{Joint Quantum Institute, NIST/University of Maryland, College Park, Maryland 20742 USA}
\affiliation{Joint Center for Quantum Information and Computer Science, NIST/University of Maryland, College Park, Maryland 20742 USA}
\affiliation{Department of Physics, Princeton University, Princeton, New Jersey 08544 USA}

\author{Andrew J. Hachtel}
\affiliation{Joint Quantum Institute, NIST/University of Maryland, College Park, Maryland 20742 USA}

\author{Marcin Kalinowski}
\affiliation{Joint Quantum Institute, NIST/University of Maryland, College Park, Maryland 20742 USA}
\affiliation{Faculty of Physics, University of Warsaw, Pasteura 5, 02-093 Warsaw, Poland}

\author{Mary E. Lyon}
\affiliation{Joint Quantum Institute, NIST/University of Maryland, College Park, Maryland 20742 USA}

\author{Alexey V. Gorshkov}
\affiliation{Joint Quantum Institute, NIST/University of Maryland, College Park, Maryland 20742 USA}
\affiliation{Joint Center for Quantum Information and Computer Science, NIST/University of Maryland, College Park, Maryland 20742 USA}

\author{Steven L. Rolston}
\affiliation{Joint Quantum Institute, NIST/University of Maryland, College Park, Maryland 20742 USA}

\author{J. V. Porto}
\affiliation{Joint Quantum Institute, NIST/University of Maryland, College Park, Maryland 20742 USA}
\email{Corresponding author: porto@umd.edu}

\date{\today}

\begin{abstract}

Long-range Rydberg interactions, in combination with electromagnetically induced transparency (EIT), give rise to strongly interacting photons where the strength, sign, and form of the interactions are widely tunable and controllable. Such control can be applied to both coherent and dissipative interactions, which provides the potential to generate novel few-photon states. 
Recently  it has been shown that Rydberg-EIT is a rare system in which three-body interactions can be as strong or stronger than two-body interactions.
In this work, we study a three-body scattering loss for Rydberg-EIT in a wide regime of single and two-photon detunings.
Our numerical simulations of the full three-body wavefunction and analytical estimates based on Fermi's Golden Rule strongly suggest that the observed features in the outgoing photonic correlations are caused by the resonant enhancement of the three-body losses.

\end{abstract}

\maketitle

Photons coherently coupled to highly excited atoms in the form of dark-state Rydberg polaritons have proven to be a versatile system for engineering strong interactions between photons. Recent experiments have shown quantum nonlinearities at the single-photon level \cite{Mohapatra2007, Dudin2012, Peyronel2012, Maxwell2013,Li2016, Paris2017, Cantu2020}, single-photon transistors \cite{Gorni2014, Tiarks2014, Gorni2016}, photonic and atomic phase gates \cite{Maller2015,Zeng2017, Levine2018, Tiarks2019}, as well as the observation of strongly-correlated photon states \cite{Firstenberg2013, Liang2018, Stiesdal2018}. 
Depending on the conditions used to generate the polaritons, the interactions can be coherent or dissipative, with controllable inherent multi-body character \cite{Gorshkov2011,Bienias2014, Moos2015, Jachymski2016,Gullans2016,Gullans2017, Gambetta2020b}. The study of few-body systems with long-range interactions can help to engineer more complex many-body quantum systems and understand their properties and potential limitations due to loss, decoherence, or recombination. Realizing precise and reliable control of three-body effects opens the door to rich phenomena, such as the universality of Efimov states \cite{Efimov1970}, the purification of a quantum gas \cite{Dogra2019}, and the emergence of strongly-correlated photonic states \cite{Liang2018, Stiesdal2018}, including fractional quantum Hall states \cite{Wojs2010, Yang2018}.

Three-body effects between Rydberg polaritons can be strong~\cite{Jachymski2016, Gullans2016, Gullans2017, Liang2018, Stiesdal2018},
which distinguishes them from the usually weak three-body forces~\cite{foot1} observed with ultracold atoms and molecules near their ground state~\cite{Buchler2007,Daley2009,Johnson2009,Mazza2010}. The three-body Rydberg polariton system has been explored experimentally for atomic clouds shorter than the range of the interactions~\cite{Stiesdal2018},  as well as  in the dispersive regime~\cite{Liang2018}.
However, the study of dissipative three-body interactions for long atomic clouds is still lacking. 

Here, we analyze tunable three-body loss of Rydberg polaritons at a high optical density, where non-perturbative effects are strong. We experimentally study and theoretically describe the tunability of the relative strength of three-body loss versus two-body loss, which is indirectly probed in the experiment by measured two- and three-photon correlation functions. 

\begin{figure}[h!]
    \centering
      \includegraphics[width=\columnwidth]{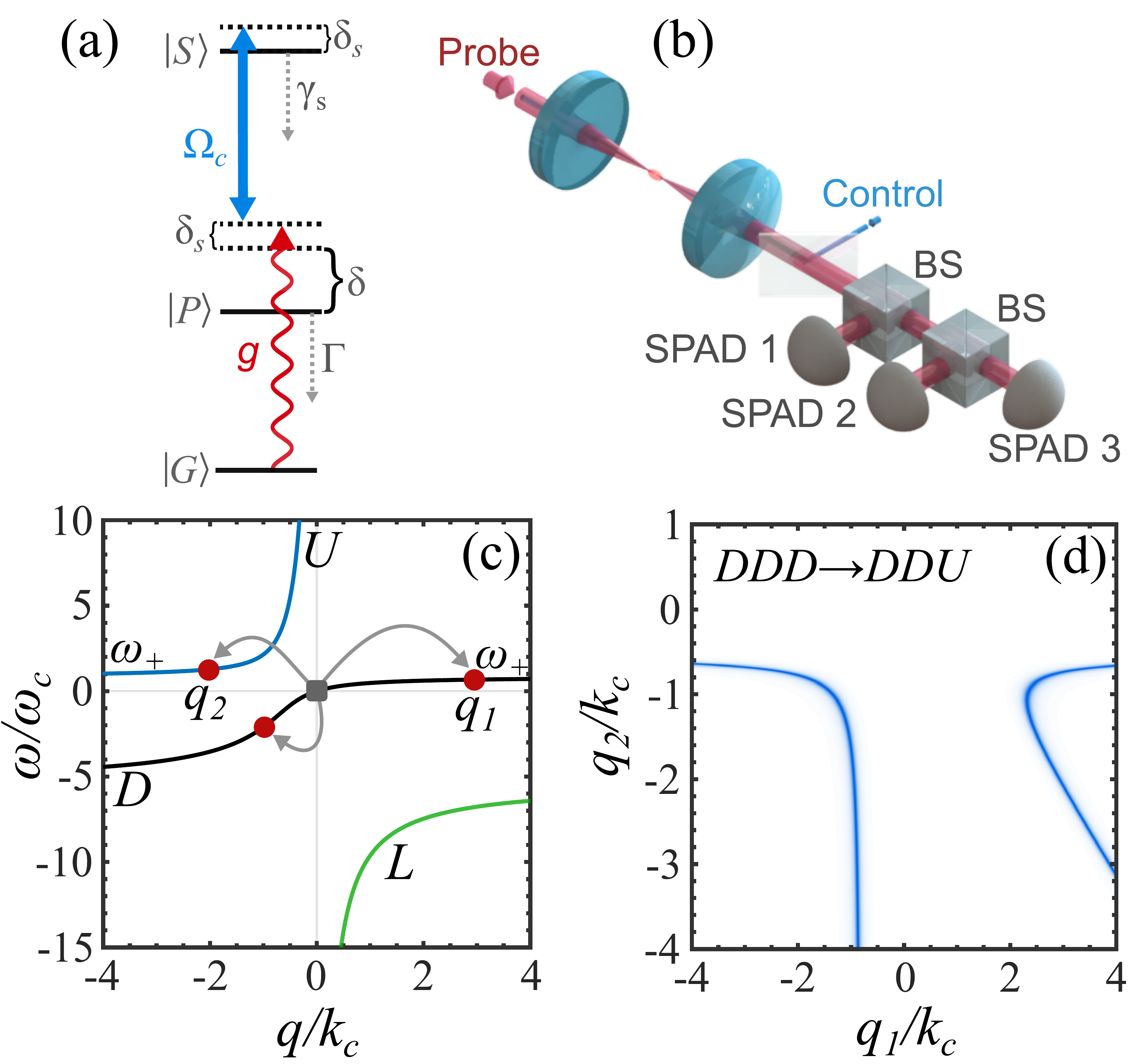}
   
    \caption{a) Atomic structure: A weak coherent probe, with collectively enhanced single-photon coupling $g$, and a strong classical field, with Rabi frequency $\Omega_c$, couples the ground state, $\ket{G}=\ket{5S_{1/2}, F=2, m_F=2}$, to the Rydberg state $\ket{S}=\ket{82S_{1/2}, m_J=1/2}$ via an intermediate state $\ket{P}=\ket{5P_{3/2}, F=3, m_F=3}$. b) Experimental setup: The probe beam and the control beam are overlapped along the propagation axis. After going through the atomic medium, the probe beam is sent to a generalized Hanbury Brown and Twiss setup to measure the second- and third-order correlation functions.
     (c) In the limit $\Gamma\ll |\delta|$, the dispersion of polaritons for experimental parameters $\delta/(2\pi)=25~\MHz$, $\delta_s/(2\pi)=0$, $\Omega_c/(2\pi)=23.5~\MHz$, for a homogeneous cloud of length $L=4.2\sigma_z$~\cite{Firstenberg2013}, with $\omega_c\equiv \Omega_c^2/4|\Delta|$ and $k_c\equiv\omega_c/v_g\approx g^2/c|\Delta|$. The black curve is the dark-state branch ($D$), while the blue and green curves are the bright states 
     ($U$ and $L$). The diagram depicts the allowed three-body loss process for three polaritons initially near the EIT resonance at $\omega_j = q_j = 0$ ($j = 1, 2, 3$ labels the three polaritons). $\omega_+$ corresponds to the energy where the dispersion for $\omega_D$ and $\omega_U$ become approximately flat. (d) Allowed final states for the three-body loss near EIT resonance. Only the process depicted in (c) is relevant for  $\deltar\approx0$. We also see that, for the plotted momenta, there is no two-body loss process allowed because there are no final states with $q_1 = 0$ or $q_2 = 0$. }
    \label{exp}
\end{figure}

The atomic-level configuration for Rydberg-EIT used to dress the incoming photon is shown in Figure~\ref{exp}(a). 
The ground state $\ket{G}$ of an ensemble of atoms is coupled to an intermediate state $\ket{P}$ by a weak quantum probe light with a collective coupling strength $g$. A strong classical field with Rabi frequency $\Omega_c$ couples $\ket{P}$ to an atomic Rydberg state $\ket{S}$. 
The Hamiltonian describing the propagation of a single excitation is given by~\cite{Fleischhauer2000,Bienias2014}
\be
H=\left( \begin{array}{c c c }
cq & g &  0\\
g& -\Delta-\delta_s & \Omega_c/2 \\
0 & \Omega_c/2 & -\Delta_s
\end{array}\right),
\label{Ham}
\ee
in the basis of $\{\mathcal{E},P,S\}$, where $\mathcal{E}$, ${P}$ and ${S}$ are the wavefunctions of the photonic component, intermediate- and Rydberg-state collective spin excitations, respectively~\cite{Fleischhauer2000} (here, we take $\hbar=1$). 
The complex detunings $\Delta=\delta+i\Gamma/2$ and $\Delta_s=\delta_s+i\gamma_s/2$ take into account the decay rates of the excited states, and $cq$ corresponds to the kinetic energy of the photon in the rotating frame, where $c$ is the speed of light. 
The rotating frame is chosen so that the incoming probe photons have zero energy. Diagonalization of Eq.~(\ref{Ham}) gives rise to three eigenstates called polaritons. For small $\delta_s$, the incoming photons propagate through the atomic medium as a hybrid photon-atom excitation with a negligible admixture of the lossy intermediate state -- so-called dark-state polaritons~\cite{Fleischhauer2002}. The coupling giving rise to the dark-state polaritons efficiently maps the strong Rydberg interaction onto the photons~\cite{Petrosyan2011, Gorshkov2011}.

The energies of the dark, $D$, bright lower, $L$, and bright upper, $U$, polaritons (depicted by $\omega_D(q)$, $\omega_L(q)$, and $\omega_U(q)$) are shown in \figref{exp}(c).
We denote the middle branch as dark $D$ because it is continuously connected to the dark state; however, for large momenta |$q$|, this branch becomes lossy.

For small $|\delta_s|\ll\omega_c$, the two-body scattering processes in which one or both of the incoming dark polaritons become lossy, are strongly suppressed~\cite{Bienias2014}, see Fig.~\ref{exp}(d) for the illustration of the $DD\rightarrow DU$ suppression. Intuitively, this suppression comes from the fact that, for $\delta_s=0$, the incoming $q=0$ polaritons are not allowed to scatter to any bright channel due to energy and momentum conservation.

\begin{figure}[ht!]
    \centering
  \includegraphics[width=\columnwidth]{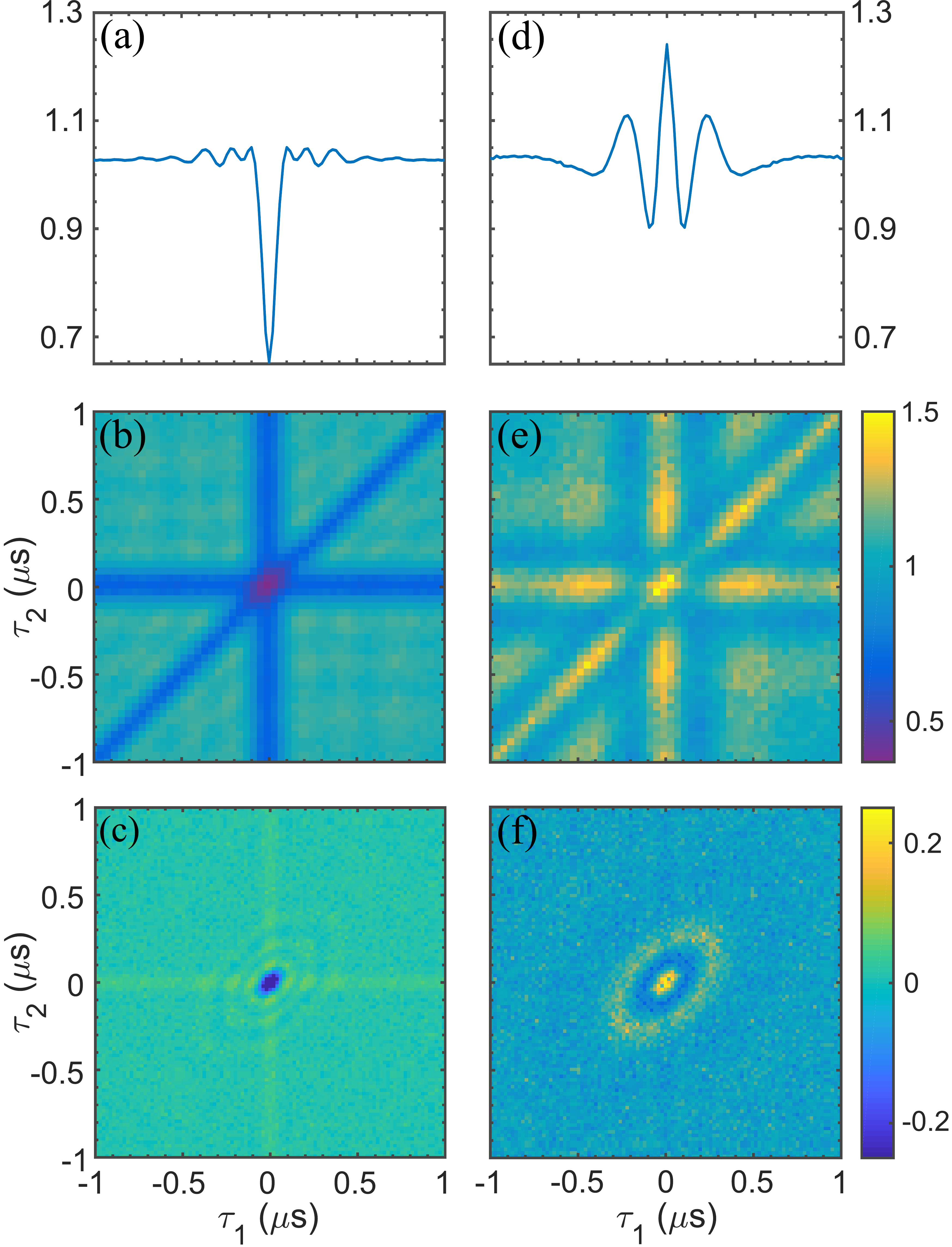}
    \caption{(a-c) Measured (a) $g^{(2)}(\tau)$, (b) $g^{(3)}(\tau_1, \tau_2)$, and (c) $\eta_3(\tau_1, \tau_2)$ for the experimental parameters indicated in the text with $\delta/(2\pi)=15~\MHz$ and $\delta_s/(2\pi)=-2~\MHz$, where we observe $\eta_3(0, 0)<0$. (d-f) Measured (d) $g^{(2)}(\tau)$, (e), $g^{(3)}(\tau_1, \tau_2)$, and (f) $\eta_3(\tau_1, \tau_2)$ for $\delta/(2\pi)=22.5~\MHz$ and $\delta_s/(2\pi)=2~\MHz$, where we observe  $\eta_3(0, 0)>0$.
    }
    \label{fig:g3_eta}
\end{figure}

However, for three photons, the scattering to lossy branches is allowed by conservation laws, which can lead to tunable three-body losses. The interplay of the shape of the interactions and of the dispersion relation can lead to resonant enhancement of three-body loss. Both the interaction potential and the dispersion relation can be tuned using $\Omega_c,\delta$, and $\delta_s$, which we explore both experimentally and theoretically in this paper.

We generate Rydberg polaritons in a cold, trapped cloud of $^{87}$Rb atoms using the three states $\ket{G}= \ket{5S_{1/2}, F=2, m_F=2}$, $\ket{P}= \ket{5P_{3/2}, F=3, m_F=3}$ and $\ket{S} =\ket{82S_{1/2}, m_J=1/2}$ (see supplemental~\cite{supplement3bLossExp} for additional details).
The weak probe beam addressing the $\ket{G}$-$\ket{P}$ transition has a beam waist of 3.3~$\mu$m and coupling strength $g/(2\pi)\simeq 10^3$~MHz. The probe waist is smaller than the Rydberg blockade radius (defined below), resulting in an effectively 1D system.
The average incoming photon rate is $R_{in}\simeq 3 \, \mu$s$^{-1}$, so that we can neglect the likelihood of more than three photons in the cloud.

The strong control beam coupling $\ket{P}$ to $\ket{S}$ is counter-propagating to the probe [Fig.~\ref{exp}(b)] and has a waist of 19~$\mu$m and a Rabi frequency of $\Omega_c/(2 \pi)=23.5\pm1.5$~MHz. 

The optically trapped atomic cloud 
consists of $\simeq10^5$ atoms cooled to 10~$\mu$K  and has an rms length of
$\sigma_r=20\pm2$~$\mu$m ($\sigma_{z}=40\pm4$~$\mu$m) in the radial (axial) direction. The resulting optical depth is $\text{OD}=37\pm4$, and we measure the linewidths to be $\Gamma/(2\pi)=7\pm1$~MHz and $\gamma_s/(2\pi)=0.4\pm0.1$~MHz. 

The impact of interactions among $n$ polaritons can be characterized by the $n$-photon correlation functions, $g^{(2)}(\tau)$ and $g^{(3)}(\tau_1, \tau_2)$ for $n=2$ and $n=3$, respectively. We measure these correlations by detecting the relative temporal delay of transmitted photons using three single-photon avalanche photodetectors (SPAD) arranged in a generalized Hanbury Brown and Twiss setup, as shown in Fig.~\ref{exp}(b). To characterize the impact of three-body 
loss relative to two-body effects at low photon rates, we use a connected correlation~\cite{Jachymski2016, Stiesdal2018}, defined as 
\begin{equation}
  \eta_3(\tau_1, \tau_2)=g^{(2)}(\tau_1)+g^{(2)}(\tau_2)+g^{(2)}(\tau_2-\tau_1)-g^{(3)}(\tau_1,\tau_2)-2.
  \label{eq:eta}
\end{equation}
In the case of dominant two-body loss, one has $\eta_3(0,0)<0$, because there is a high probability of absorbing at least one out of two or three incoming photons and both $g^{(2)}(0)$ and $g^{(3)}(0,0)$ are suppressed (strong two-body repulsion \cite{Cantu2020} has a similar effect).

On the other hand, if two-body loss is small and dispersive two-body interactions are weak or attractive such that $g^{(2)}(0) \ge  1$, while three-body loss is strong such that  $g^{(3)}(0,0) < 1$, we get $\eta_3(0,0)>0$. 
Therefore, we use a positive value of $\eta_3(0,0)$ as a signature of strong three-body losses in the system. Figure~\ref{fig:g3_eta} shows the measured second-order, third-order, and connected third-order correlation functions for two parameter choices corresponding to  $\eta_3(0,0)<0$ [Fig.\ \ref{fig:g3_eta}(a-c)] and $\eta_3(0,0)>0$ [Fig.\ \ref{fig:g3_eta}(d-f)].  

\begin{figure}[]
    \centering
  \includegraphics[width=\columnwidth]{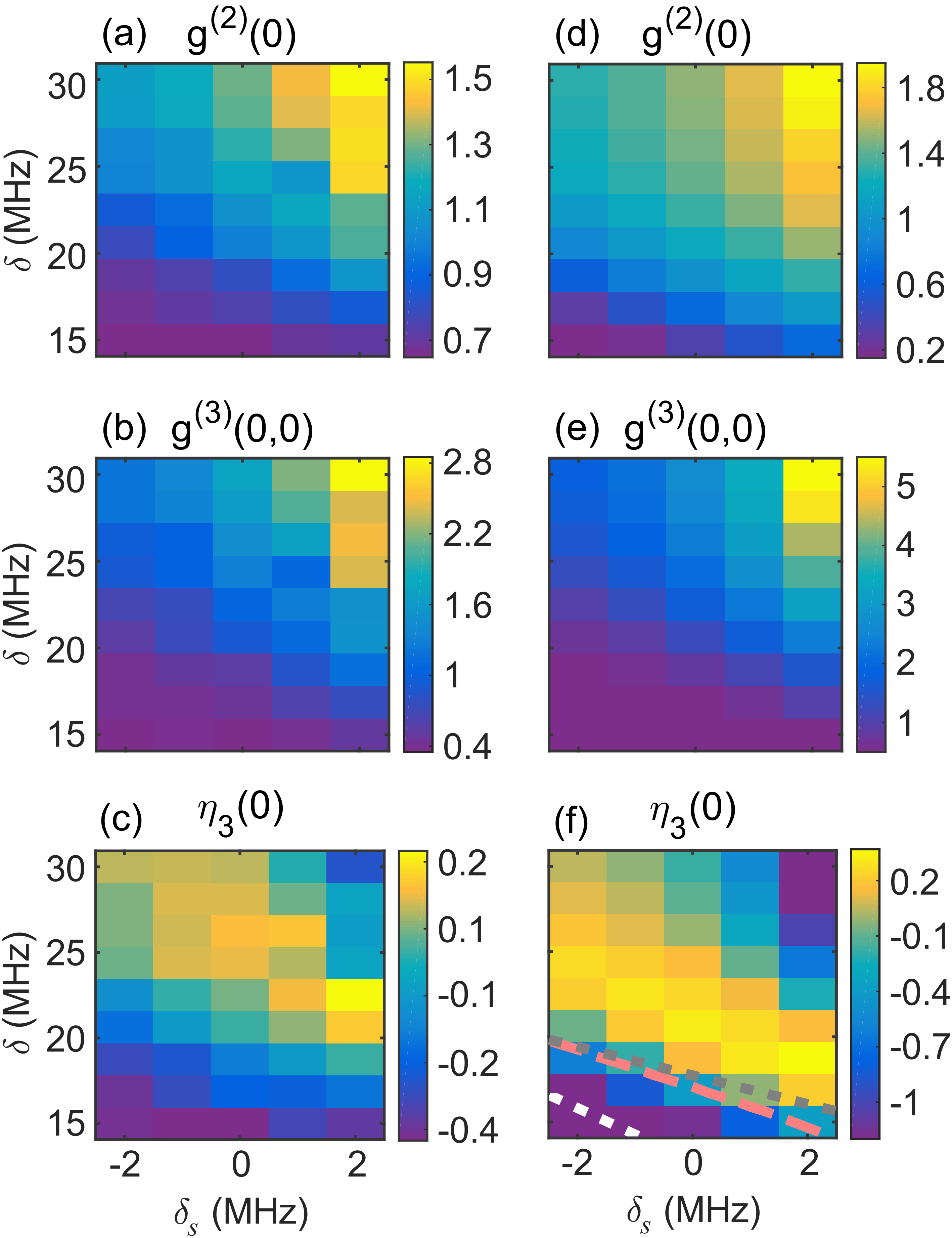}
    \caption{(a)-(c) Experimental data of the second-order, $g^{(2)}(0)$, third-order, $g^{(3)}(0,0)$, and connected third-order, $\eta_3(0,0)$, correlation functions with $\Omega_c/(2\pi)=23.5\pm1.5$~MHz, for a cloud with OD=37$\pm4$ and $\sigma_z=42\pm4$~$\mu$m. (d)-(f) Numerical simulations for the same correlation functions. Parameters used for the simulations are: OD = 37, $\Omega_c/2\pi=25$ MHz, $\Gamma/2\pi=7$ MHz, $\gamma/2\pi=0.3$ MHz and $\sigma_z=40~\mu$m. Regions with $\eta_3(0,0)>0$ indicate excess of three-body loss with respect to two-body loss. The dashed lines indicate the enhanced three-body loss predicted by the Fermi's Golden Rule calculation (see the text).} 
    \label{fig:corr}
\end{figure}

We  show the measured correlation functions, $g^{(2)}(0)$, $g^{(3)}(0,0)$,  and $\eta_3(0,0)$, as a function of $\delta$ and $\delta_s$, at fixed $\Omega_c$, in Figure~\ref{fig:corr}(a-c).
The region where $\eta_3(0,0)>0$ (indicative of dominant three-body loss) occurs in a roughly linear band in $\delta$-$\delta_s$ space with a negative slope. Figure~\ref{fig:corr}(d-f) shows $g^{(2)}(0)$, $g^{(3)}(0,0)$, and $\eta_{3}(0,0)$ obtained by numerically solving the Schr\"{o}dinger equation for the two- and three-polariton wavefunctions propagating through the Rydberg-EIT medium~\cite{Gullans2013} using similar parameters to the experimental values. 
We find good qualitative agreement between the numerical calculation and experiment: we reproduce the antibunching to bunching behavior in $g^{(2)}(0)$ and $g^{(3)}(0,0)$, as well as the resonant-like feature of three-body loss in $\eta_3(0, 0)$ observed in the experimental data. 

To understand the source of the resonant-like enhancement of $\eta_3(0,0)$, we first discuss the form of two-body interactions, then describe the three-body scattering process using a Fermi's Golden Rule approximation, 
and finally compare the results with the experimental and numerical observations of the resonant three-body loss feature (\figref{fig:corr}).

Atoms in Rydberg states interact via van der Waals interactions $V(r)=C_6/r^6$.   
The effective interaction between two dark-state polaritons is given by~\cite{Bienias2014}
\beq
\Ve(\omega,r)=\frac{V(r)}{1-\chibar(\omega) V(r)}.
\eeq
Here, $\bar{\chi}$ characterizes the saturation 
of the potential at distances less than the blockade radius $r_b=(C_6|\bar{\chi}|)^{1/6}$~\cite{Bienias2014} and is given by

\beq
\bar{\chi}(\omega)=\frac{-\Omega _c^2+4 \tilde{\Delta} ^2+6 \tilde{\Delta}  \nu +2 \nu ^2}{2 (\tilde{\Delta} +\nu ) \left(\nu  (2 \tilde{\Delta} +\nu )-\Omega _c^2\right)}
\eeq

with $\nu=\omega+2\Delta_s$ and $\tilde{\Delta}=\delta+i\Gamma/2-i\gamma_s/2$.
Note that $\gamma_s\ll \Gamma$, therefore, we can neglect the difference between $\tilde{\Delta}$ and $\Delta=\delta+i\Gamma/2$, which we do in the analytical expressions that follow. 
In our experiment, $r_b$ ranges from 7~$\mu$m to 10~$\mu$m.

Using $V_e$, we analyze the three-body scattering rate $\beta$, for incoming dark-state polaritons 
near EIT resonance due to processes like the one indicated in Fig.~\ref{exp}(c).
We perform our analysis in the ideal limit of zero dissipation and then analytically continue to finite $\Gamma$ and $\gamma_s$.

\begin{figure}[]
    \centering
  \includegraphics[width=\columnwidth]{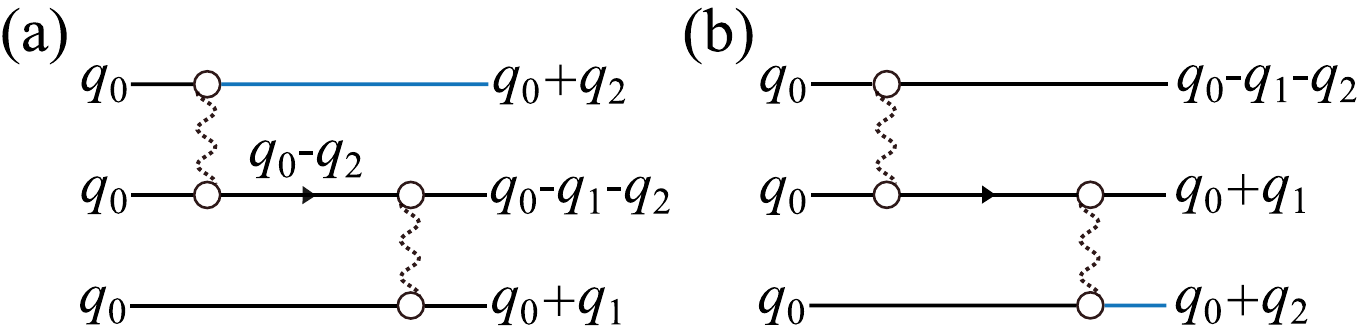}
    \caption{ 
    Lowest-order diagrams that contribute to three-body loss. We replace the bare interaction by the effective two-body interaction potential and take all external lines to be on-shell in evaluating these diagrams. The black lines indicate polaritons in the dark branch, and the blue lines indicate polaritons scattered to the upper-bright branch. We use the full propagator for the $S$-states in the virtual state (depicted by the arrowed line), which includes contributions from all three branches. In addition to this diagram, there are five similar diagrams (total of six) for both (a-b) obtained by permuting inputs and permuting outputs.
    }
    \label{theoDiagram}
\end{figure}

The lowest-order diagrams contributing to the scattering rate $\beta$ are second-order in $V_e$. 
The conservation of energy and momentum puts additional restrictions on the available open scattering channels.
In \figref{theoDiagram}(a-b), we show the leading contributions to $\beta$ which involve scattering to $\mid \mid U$ with $\mid$ gaining large $q$ and thus also becoming lossy. 
Other allowed processes, such as scattering to $ D  U L$ %$D U U $
can be neglected due to the weaker effective interactions involving these bright polaritons because of their small Rydberg amplitude. 

The incoming polaritons have $ \omega_D(q_0) = 0$. In general, the incoming momentum $q_0\neq 0$ for $\delta_s\neq 0$, but, for brevity of presentation, we show the expressions for $\delta_s=0$ and $q_0=0$. Within a Fermi's Golden Rule calculation, the diagrams in Figs.~\ref{theoDiagram}(a) and (b) contribute, respectively, the first and second terms inside the absolute value in the following expression for $\beta$:
\beqa
\beta\, &=& \frac{18}{\pi}\integral{q_1}\text{d}q_2 |S_D^{0}|^6  |S_D^{q_1}|^2 |S_D^{-q_1-q_2}|^2 |S_U^{q_2}|^2\nn \\
\!\!\!&\times&\,{\Big |}
\VoneD_{q_2}[0] G_{ss}[-q_2,-\omega_U(q_2)]\VoneD_{q_1}[-\omega_U(q_2)]\nn\\
\!\!\!&+&\!\VoneD_{q_1+q_2}[0] G_{ss}[q_1\!+\!q_2,\!-\!\omega_D(\!-\!q_1\!-\!q_2)]  \VoneD_{q_2}[\!-\!\omega_D(-q_1\!-\!q_2)]{\Big |}^2\nn\\
\!\!\!&\times&\delta\left(\omega_U(q_2)+\omega_D(q_1)+\omega_D(-q_1-q_2)  \right).
\label{betaDef}
\eeqa
Here, $\VoneD_q[\omega]$ is the Fourier transform of $V_e(\omega,r)$, $G_{ss}$ is the single-body propagator projected onto the Rydberg state, $\omega_U(q)$ is the dispersion for the upper-bright branch, and $S^q_\nu$ is the overlap of the Rydberg state with a polariton at momentum $q$ on branch $\nu\in\{D, U\}$ (see supplemental~\cite{supplement3bLossExp} for additional details). 

The behavior of Eq.~(\ref{betaDef}) depends on the interaction strength, which can be quantified by $\varphi= |r_b/\sqrt{\bar{\chi}/m}|$, where $m=-2g^4/\Delta\Omega^2c^2$.  
For $|\delta| \gg \Omega_c/2$, $\varphi$ simplifies to $\OD_b\Gamma/4|\Delta|$ (which, up to a constant factor, is the phase a stationary Rydberg excitation would imprint on a passing polariton~\cite{Gorshkov2011}), where we use $\OD_b=\text{OD}  r_b /{\sqrt{2\pi}\sigma_z}$, 
which is the optical depth per blockade radius corresponding to the maximal density of a Gaussian cloud with rms $\sigma_z$. In our experiment, $\OD_b$ 
is < 4, thus for the detunings considered here $\varphi <0.3$.

In the moderately interacting regime of $\varphi < 1$, which applies to our experiment, we can simplify the evaluation of Eq.~(\ref{betaDef}) by noting that the dispersions for $\omega_D$ and $\omega_U$ become approximately flat and saturate to $\omega_+$ (see \cite{supplement3bLossExp}) in the relevant range of the momentum transfer $\sim 1/r_b$ being larger than the characteristic threshold momentum $k_c$ [see Fig.~\ref{exp}(c,d)]. 

This results in the second term in Eq.~(\ref{betaDef}) vanishing because $\VoneD_{q_2}[-\omega_D(-q_1-q_2)]\approx\VoneD_{q_2}[2\omega_+ ] \ra 0$, so $\beta$ 
simplifies to~\cite{foot2} 
\beq
\frac{18}{\pi}\integral{q}\frac{1}{v_g(-2\omega_+)}\left| \VoneD_q[0]
G_{ss}[q_2\ra\infty,-\omega_+]
\VoneD_q[-\omega_+]
\right|^2,
\label{beta}
\eeq
which has a complicated dependence on the experimental parameters. We concentrate on qualitative features of Eq.~(\ref{beta}) to understand the behavior of $\beta$. In the regime where $\Omega_c\ll\ | \delta|$, the scattering rate can be simplified to
$ \beta\propto \varphi r_b^2 \,{\Omega_c^2}/{\delta}
$.
Here, $\beta$ increases with increasing interaction $\varphi$, but does not feature any resonances as a function of $\delta$.

In contrast, for $\Omega_c\sim\delta$, the scattering rate given by Eq.~(\ref{beta}) could have resonant behavior for two reasons. First, the density of outgoing states, characterized by $1/v_g(-2\omega_+)$, could diverge as a function of $\delta$. 
Second, the interaction vertices $\VoneD_q[0]$ or $\VoneD_q[-\omega_+]$, which are inversely proportional to ${\bar{\chi}(0)}$ and ${\bar{\chi}(-\omega_+)}$, could have a resonance due to the vanishing value of $\chibar$. 
This divergence in the interaction vertices will be smoothed out in our regime of finite $\Gamma,\gamma_s$, but will still have a significant impact on $\beta$.
We find that the divergence in the density of states is nearly canceled by the simultaneous vanishing of $\VoneD_q[-\omega_+]$ (see supplemental~\cite{supplement3bLossExp} for details) and the density of states, therefore, does not contribute to the resonant behavior. 

The interaction vertices $\VoneD_q[0]$ and $\VoneD_q[-\omega_+]$ diverge for $\delta$ approaching specific respective detunings $\delta_{0}$ and $\delta_{\rs +}$ where 
$\bar{\chi}(0)$ and $\bar{\chi}(-\omega_+)$ respectively 
vanish. In the experimentally relevant limit
$|\delta_s|\ll \Omega_c,|\Delta|$, the expressions for $\delta_{0}$ and $\delta_{+}$ simplify to
$
\delta_{0} = \frac 1 2 \Omega_c-\frac 3 2 \delta_s$, $
\delta_{\rs +}=\frac{1}{2} \sqrt{\frac{1}{6} \left(\sqrt{33}+6\right)}\Omega_c-\frac{1}{264} \left(\sqrt{33}+209\right) \delta_s\approx
0.7 \Omega _c-0.8 \delta_s 
$.
These dependencies are shown in \figref{fig:corr}(f): the gray dotted line depicts $\delta_{\rs +}$, whereas the white dotted line depicts $\delta_{0}$.
In our system, the decay $\Gamma$ leads to such significant broadening of the two resonances that the two peaks are no longer distinguishable, leading to a single, effective resonant feature for $\beta$. 
In \figref{fig:corr}(f), the pink dashed curve depicts the value  of $\delta$ for which $|\beta|$ given by Eq.\ (\ref{beta}) is maximal for a fixed $\delta_s$. 

The maximal curve is closer to the $\delta_{\rs +}$ line because, for our parameters, this resonance is stronger than the $\delta_{0}$ resonance. The overall resulting resonance is a three-body effect because it predominantly comes from the $\delta_{\rs +}$ resonance, which is not present for the two-body scattering. 
In the vicinity of a divergent $1/\bar{\chi}$, the interaction strength could become large and negative, leading to the emergence of the second bound state, which would happen for $\varphi\approx 3$~\cite{Bienias2014}. 
This nonperturbative effect could hinder the applicability of the \FGR.
However, since in our system $\varphi < 0.3$ (due to dissipation), we
neglect the second bound state.

The three-body scattering effects measured in this experiment probe nonperturbative processes, even in the moderately interacting regime $\varphi < 1$. The \FGR calculation can be considered as a perturbative and incomplete description of the three-body scattering.  Nonperturbative effects can be more accurately captured by introducing an effective three-body interaction 
between dark-state polaritons~\cite{Jachymski2016,Gullans2016,Liang2018}.  
These $N$-body interactions, however, are  a momentum and frequency dependent quantity in free space, whose full description requires the exact solution to the $N$-body problem. Important steps in developing an approximate, consistent renormalization group treatment of three-body forces have recently been made by carefully analyzing single-mode-cavity setups~\cite{Kalinowski2019Prep}.

{\it Summary \& Outlook---}
In summary, we demonstrate the ability to tune Rydberg-polariton interactions leading to three-body losses. 
These interactions are analyzed using the few-body auto-correlation functions of the outgoing field. Our numerical simulations reproduce the experimentally observed features with good qualitative agreement.
We provide a physical description of the tunable losses based on a \FGR treatment of the scattering process of three dark-state polaritons to two lossy dark-state polaritons and a bright-state polariton.
It should be possible to observe a sharper and stronger three-body loss resonance by detuning further from single-photon resonance to decrease the dissipation from the intermediate state.  The resonant regime requires $\Omega_c\sim\delta$, so the optical power needed to achieve strong enough Rabi frequencies can be experimentally challenging. 
Pushing further into this regime in similar experimental setups would enable the production of a novel three-photon number filter.

Our work is a demonstration of the tunability offered by Rydberg systems, showing promising directions in the study and control of few and many-body physics of strongly interacting photons. 
For example, increasing the strength of the interactions $C_6$ and the optical density gives the ability to tune the scattering length and observe another bound state associated with a photon-photon scattering resonance \cite{Bienias2014}.
Extending the system to three dimensions and altering the polariton effective mass and interactions, along the transverse and axial directions,  could result in photonic Efimov trimers \cite{Gullans2017}. Another interesting theoretical and experimental direction involves studying unconventional topological and spin-liquid phases with three-body forces, especially in the two-dimensional geometry~\cite{Jia2018}.

\begin{acknowledgments}

We thank Hans Peter B\"{u}chler, Yidan Wang, and Elizabeth Goldschmidt
for insightful discussion. 
We are also grateful to Nathan Fredman for his constributions to the experiment.
P.B., M.K., A.C, D.O.-H, S.L.R., J.V.P., and A.V.G. acknowledge support from the United States Army Research Lab’s Center for Distributed Quantum Information (CDQI) at the University of Maryland and the Army Research Lab, and support from the National Science Foundation Physics Frontier Center at the Joint Quantum Institute (Grant No. PHY1430094).  P.B., M.K., and A.V.G. additionally acknowledge support from AFOSR, ARO MURI, AFOSR MURI, DoE ASCR Quantum Testbed Pathfinder program (award No. DE-SC0019040), U.S. Department of Energy Award No. DE-SC0019449, DoE ASCR Accelerated Research in Quantum Computing program (award No. DE-SC0020312), and NSF PFCQC program. 
\end{acknowledgments}

\clearpage
\onecolumngrid

\begin{center}

\newcommand{\beginsupplement}{%
        \setcounter{table}{0}
        \renewcommand{\thetable}{S\arabic{table}}%
        \setcounter{figure}{0}
        \renewcommand{\thefigure}{S\arabic{figure}}%
     }
%\beginsupplement

\textbf{\large \section{Supplemental Material: Tunable three-body loss in a Rydberg nonlinear medium}}
\end{center}

\newcommand{\beginsupplement}{%
        \setcounter{table}{0}
        \renewcommand{\thetable}{S\arabic{table}}%
        \setcounter{figure}{0}
        \renewcommand{\thefigure}{S\arabic{figure}}%
     }
%\renewcommand{\thefigure}{S\arabic{figure}}

%%%%%%%%%%% Merge with supplemental materials %%%%%%%%%%
%%%%%%%%%% Prefix a "S" to all equations, figures, tables and reset the counter %%%%%%%%%%
\setcounter{equation}{0}
\setcounter{figure}{0}
\setcounter{table}{0}   
\setcounter{page}{1}
\makeatletter
\renewcommand{\theequation}{S\arabic{equation}}
\renewcommand{\thefigure}{S\arabic{figure}}
\renewcommand{\thetable}{S\arabic{table}}
\renewcommand{\bibnumfmt}[1]{[S#1]}
\renewcommand{\citenumfont}[1]{S#1}

\section{Experimental Methods \label{sec:exp}}

\subsection{Atom preparation}

We load $^{87}$Rb atoms into a magneto-optical trap from background vapor for 250~ms. We use a $\Lambda$-gray molasses~\cite{Rosi2018} scheme on the D2 transition to cool the atoms down to 10~$\mu$K, and confine them in an optical dipole trap made of 1003-nm light. The trapped atomic cloud has dimensions $\sigma_r=20$~$\mu$m in the radial direction, and an axial extension along the probe propagation of $\sigma_{z}=40$~$\mu$m. The resulting optical depth is $\text{OD}=37\pm4$ with $\approx10^5$ atoms. The dipole trap consists of three beams that intersect at the probe beam focus. Two of the beams form a $\approx \pm11^{\circ}$ crossed trap with respect to the $z$-axis (along the probe direction), while a third elliptical shaped beam travels in the $y$-axis; all beams lie on the same ($z$-$y$) plane. We optically pump the atoms into the stretched state $\ket{5\text{S}_{1/2},F=2,m_F=2}$, using $\sigma_+$-polarized light blue-detuned from the $F=2$ to $F'=2$, D1 transition. 

\subsection{Two-photon excitation}
We couple the ground and Rydberg state with a two-photon transition. A 780-nm weak probe field addresses the transition from the ground state, $\ket{G}=\ket{5\text{S}_{1/2},F=2,m_F=2}$  to the intermediate state, $\ket{P}=\ket{5\text{P}_{3/2},F=3,m_F=3}$; a strong control field addresses the transition from the intermediate state to the Rydberg state, $\ket{S}=\ket{82\text{S}_{1/2},J=1/2,m_J=1/2}$ with a wavelength of 479 nm.

The probe and control lasers are frequency stabilized via a Pound-Drever Hall lock scheme using an ultra-low expansion (ULE) cavity with a linewidth $<10$ kHz. We use the probe light that has been transmitted and filtered by the ULE cavity to reduce phase noise during the two-photon excitation~\cite{Leseleuc2018}.

The measured linewidth for both the $\ket{G}$-$\ket{P}$ and $\ket{P}$-$\ket{S}$ transitions are $\Gamma/(2\pi)=7\pm1$~MHz and $\gamma/(2\pi)=0.4\pm0.1$~MHz, respectively. 
The latter is broadened beyond the natural linewidth by various dephasing mechanisms e.g. differential light shifts and Doppler broadening.
%, laser phase noise and ground-state Rydberg collisions.

%broadening contributions from the natural linewidth of the Rydberg state, light shifts from the dipole light, Doppler  %coupling to other nearby Rydberg states, %\todo{PB to experimentalists: what type of coupling?}
%laser phase noise from the control laser light, and ground-state Rydberg collisions.

We focus the probe beam down to a $1/e^2$ waist of $w_p\approx3.3$ $\mu$m to ensure the system is effectively  one dimensional ($w_p < r_b$), where the blockade radius $ r_b$ ranges from 7 to 10$\mu$m. The control beam is counter-propagating to the probe and focused to a beam waist of $w_c\approx19$ $\mu$m. The larger beam waist provides an approximately uniform control field across the probe area. 

We interrogate the atoms continuously for 100~ms before we repeat the loading and cooling cycle, resulting in an experimental duty-cycle of $\simeq $0.13.

\subsection{Light collection and filtering}

After exiting the chamber, the probe light passes through a polarization beam splitter (PBS), and a narrow 1-nm spectral filter (Alluxa 780-1 OD6~\cite{commercial}) to reduce the amount of background light reaching the detectors, due to e.g. leakage of room light and broad-band fluorescence from the control laser.
%to reduce broad-band fluorescence from the chamber windows and leakage light that goes into the detection system. 
The probe is equally split in three using an arrangement of half-waveplates and two PBS, and is then coupled to multi-mode fibers and sent to single-photon avalanche photo detectors (SPAD) Excelitas SPCM-780-13~~\cite{commercial}).

\subsection{Correlation measurements}

The detection events from the SPADs are recorded as time-stamps by a triggered time-tagger device (Roithner TTM8000~~\cite{commercial}). We calculate the second-order correlation function as, %we count the coincidence events $n_{12}$ between each detector pair as a function of the relative time of detection (where the time-tagger is triggered to an absolute time), then we normalize the correlation function with the total number of recorded events $n_i$ by each individual detector, $g^{(2)}=\frac{ n_{12}(\tau)}{\langle n_1\rangle\langle n_2\rangle}$. 
\be
g^{(2)}(\tau)=\frac{ N_{12}(\tau)}{ \frac{1}{8}\sum_{m=1}^4N_{12}(\tau\pm m T)},
\ee
where $N_{12}(\tau)$ are the measured coincidences as a function of the relative time $\tau$ of detection between each SPAD pair accumulated for 1500 experimental cycles. The normalization is done by counting the coincidence events for $\tau+mT$, where $T=100~\mu$s and $m$ is an integer number. The coincidences for large $\tau$ are given by $N_{12}(\tau\pm m T)=N_1 N_2 \Delta \tau/T\rs{exp}$. Here, $\Delta \tau$ is the binning time (which corresponds to 20~ns in the experimental data shown in the main text), and $N_j$ is the average detection counts for the $j$-th SPAD in a time-bin $\Delta\tau$, and $T\rs{exp}$=100~ms is the total experimental time. This normalization allows to cancel slow experimental drifts with time scales longer than $T$, that could be imprinted in the correlation measurements.

The third order correlation function is calculated as,
\be
g^{(3)}(\tau_1, \tau_2)=\frac{ N_{123}(\tau_1, \tau_2)}{ \frac{1}{56}\sum_{m=1, n=1, m\neq n}^4 N_{123}(\tau_1\pm m T, \tau_2\pm n T)},
\ee
where $N_{123}(\tau_1,\tau_2)$ are the measured coincidences as a function of two relative times $\tau_1$ and $\tau_2$, and is given by $N_{123}(\tau_1,\tau_2)=N_1 N_2 N_3 (\Delta \tau/T)^2$. For the normalization of $g^{(3)}(\tau_1, \tau_2)$, we ignore the indexes where $m=n$, since $g^{(3)}(\tau, \tau)=g^{(2)}(0)$.

%Similarly, for the third-order correlation function, we measure the coincidence $n_{123}$ between all three detectors for two relative times, $g^{(3)}=\frac{\langle n_{123}(\tau_1, \tau_2)\rangle}{\langle n_1\rangle\langle n_2\rangle\langle n_3\rangle}$. Here, is over 1,500 experimental cycles for each data point in Fig.~2 in the main text. 

\section{Fermi's Golden Rule Calculation \label{sec:theory}}

Here, we present the \FGR calculation described in the main text.  For  Rydberg polaritons, only interactions between Rydberg levels are non-negligible. Therefore, in the basis of atomic and photonic states, only the $T$-matrix between Rydberg states is non-zero.

\begin{figure}[h] 
\includegraphics[width=.4\columnwidth]{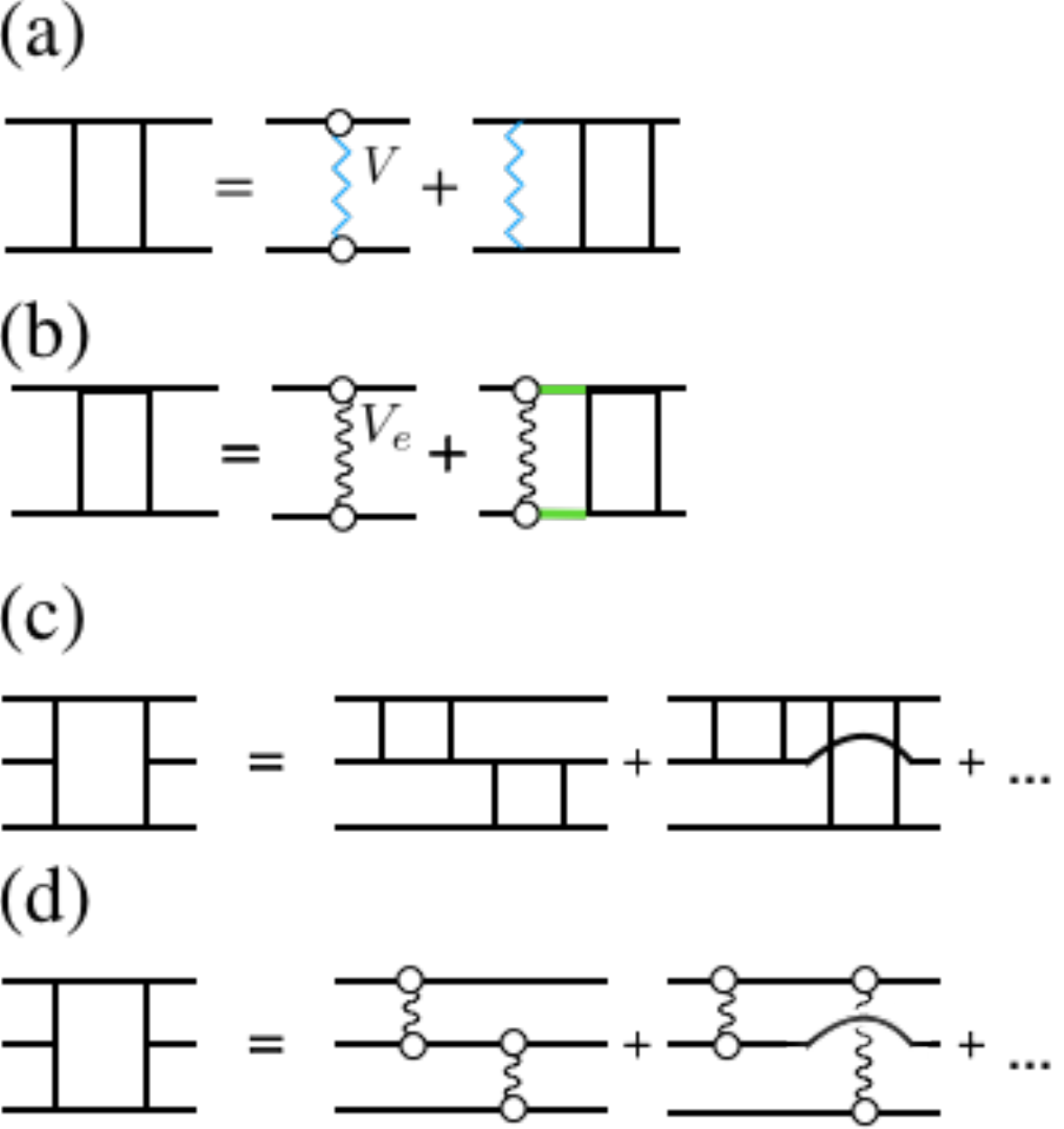}
\caption{$T$-matrix representation of the two-body (a-b) and three-body (c-d) problem. Wiggly black line depicts $V_e$, whereas polygonal blue line depicts $V$. See main text for the details.}
\label{diagrams}
\end{figure}

The Lippmann-Schwinger equation  for the two-body $T$-matrix is shown in \figref{diagrams}(a), where in all diagrams henceforth the  rectangles denote two- and three-body $T$-matrices.

From Ref.~\cite{SBienias2014}, we know that the $T$-matrix equation for the two-body scattering problem can be rewritten using effective interactions $V_e$, see \figref{diagrams}(b).  This approach has the advantage that effective interaction is non-divergent in the relative distance between the polaritons, in contrast to the initial $T$-matrix equation written in terms of $V$. Note that, in this case,  the two-body propagator (depicted as two green  horizontal lines between $V_e$ and $T$) includes only the poles that have a nontrivial momentum dependence (the momentum-independent spin-wave contribution $\bar{\chi}$ to the two-body propagator $\chi$ is taken into account in $V_e$).

The equations for the three-body $T$-matrix are called Faddeev equations. The three-body $T$-matrix is written as an infinite sum of diagrams involving two-body $T$-matrices,  \figref{diagrams}(c), in a mathematically consistent way.

To estimate $\beta$, for $\varphi<1$, we can consider only diagrams contributing in the leading second order in $V_e\sim\varphi$, see  \figref{diagrams}(d). In other words, we replace the two-body $T$-matrix by the first term $V_e$ on the right-hand side of the two-body Lippman-Schwinger equation in \figref{diagrams}(b).

For the calculation of $\beta$, we are interested in the $T$-matrix elements between incoming dark polaritons and outgoing lossy polaritons. For this purpose, it is useful to consider the $T$-matrix in the polaritonic  basis; and, in the expressions for $\beta$, we  include only the connected diagrams with the additional condition that they do not end in the three dark polaritons.
Finally, as explained in the main text, we find that the leading contribution to the \FGR scattering rate $\beta$ comes from the scattering to the $DDU$ channel--the corresponding diagrams are shown in Fig.~4 of the main text.

%\paragraph
%\paragraph{Additional analytical expressions.---}
\subsection{Additional analytical expressions}
Here we present expressions for the single-body propagator and for $\omega_+$, as well the expressions used to justify that the resonance in the \FGR integral caused by the density of states can be neglected. 

%modification on 20200630
For the sake of brevity, all the expressions in this subsection are presented for  $\gamma_s\ll\Gamma$. In order to arrive at more general expressions, one should replace $\Delta$ by $\tilde{\Delta}=\Delta-i\gamma_s/2$.
% end of modification

The single-body propagator $G_{ss}[k,\omega]$, used in the main text, is given by
%based on sec:def in reproducing_MJG_estimate_of_v7.nb
\beq
\frac{(\omega -c k) \left(\Delta +\Delta _s+\omega \right)-g^2}{\left(\Delta _s+\omega \right) \left((\omega -c k) \left(\Delta +\Delta _s+\omega \right)-g^2\right)+\frac{1}{4} \Omega ^2 (c k-\omega )}.
\eeq
while $\omega_+$ is given by
\beq
\omega_+=\frac{1}{2} \left(-\Delta +\sqrt{\Delta ^2+2 \Delta  \Delta _s+\Delta _s^2+\Omega ^2}+\Delta _s\right).
\eeq
\begin{widetext}

%\subsection
%\paragraph
Next we discuss {the cancellation of the resonance from the density of states with a zero from $1/\bar{\chi}$ within the \FGR calculation. }
Expression for $1/\bar{\chi}[-\omega_+]$ takes the form
\beq
\frac{\left(-3 \Delta +\sqrt{\left(\Delta +\Delta _s\right){}^2+\Omega ^2}-3 \Delta _s\right) \left(-3 \Delta ^2+3 \Delta  \sqrt{\left(\Delta +\Delta _s\right){}^2+\Omega ^2}+3 \Delta _s \sqrt{\left(\Delta +\Delta _s\right){}^2+\Omega ^2}-10 \Delta  \Delta _s-5 \Delta _s^2+\frac{3 \Omega ^2}{2}\right)}{2 \left(-8 \Delta ^2+4 \Delta  \sqrt{\left(\Delta +\Delta _s\right){}^2+\Omega ^2}+3 \Delta _s \sqrt{\left(\Delta +\Delta _s\right){}^2+\Omega ^2}-13 \Delta  \Delta _s-5 \Delta _s^2+\frac{\Omega ^2}{2}\right)},
\eeq
which, for $\Delta_s\ra 0$, equals
\beq
\frac{\left(\sqrt{\Delta ^2+\Omega ^2}-3 \Delta \right) \left(3 \Delta  \sqrt{\Delta ^2+\Omega ^2}-3 \Delta ^2+\frac{3 \Omega ^2}{2}\right)}{2 \left(4 \Delta  \sqrt{\Delta ^2+\Omega ^2}-8 \Delta ^2+\frac{\Omega ^2}{2}\right)}.
\label{chiResonance}
\eeq
The density of states is proportional to $ 1/v_g(-2\omega_+)$, which, for $g\gg\Omega,|\Delta_s|,|\Delta|$,  is equal to 
\beq
\frac{g^2 \left(\Delta -\sqrt{\Delta ^2+2 \Delta  \Delta _s+\Delta _s^2+\Omega ^2}\right){}^2+\frac{g^2 \Omega ^2}{4}}{c \left(3 \Delta ^2-3 \Delta  \sqrt{\Delta ^2+2 \Delta  \Delta _s+\Delta _s^2+\Omega ^2}+2 \Delta  \Delta _s+\Delta _s^2+\frac{3 \Omega ^2}{4}\right){}^2},
\eeq
which, for $\Delta_s\ra 0$, in turn equals
\beq
\frac{g^2 \left(-2 \Delta  \sqrt{\Delta ^2+\Omega ^2}+2 \Delta ^2+\frac{5 \Omega ^2}{4}\right)}{9 c \left(-\Delta  \sqrt{\Delta ^2+\Omega ^2}+\Delta ^2+\frac{\Omega ^2}{4}\right)^2}.
\label{resonance}
\eeq
The divergence from the denominator of Eq.\ (\ref{resonance})  at $\Delta\approx \Omega/2\sqrt{2}$  cancels with the square of  $\left(\sqrt{\Delta ^2+\Omega ^2}-3 \Delta \right)$ from the numerator of Eq.\ (\ref{chiResonance}). For nonzero $\Delta_s$, this cancellation is approximate. Furthermore, the dependence of $r_b$ on  $\bar{\chi}$ also leads to a residual divergence. Both effects, however, are not significant due the non-negligible imaginary component of the detunings.
%The divergence from the denominator of \eqref{resonance}  at $\Delta\approx \Omega/2\sqrt{2}$  cancels with the part $\left(\sqrt{\Delta ^2+\Omega ^2}-3 \Delta \right)$ of numerator squared from \eqref{chiResonance}. For nonzero $\Delta_s$ this cancellation is approximate. We note that additionally $r_b$ dependence on the $\bar{\chi}$ leads to the residual divergence. Both effects however are not significant due the non-negligible imaginary component of the detunings.
\end{widetext}


\begin{thebibliography}{10}
\bibitem{foot2}
Where the relevant $|S_{\nu}^q|$ are $\sim 1$, and do not change the position
  of the discussed resonances.
\end{thebibliography}
\begin{thebibliography}{10}
\bibitem{commercial}
The identification of commercial products in this paper does not imply
  recommendation or endorsement by the National Institute of Standards and
  Technology nor does it imply that the items identified are necessarily the
  best available for the purpose.
\end{thebibliography}
\end{document}